\documentclass[%
superscriptaddress,
preprint,
showpacs,preprintnumbers,
 amsmath,amssymb,
 aps,
prstab,
]{revtex4-1}

\usepackage{graphicx,epsfig}% Include figure files
\usepackage{dcolumn}% Align table columns on decimal point
\usepackage{bm}% bold math
\usepackage[colorlinks=false,linkcolor=blue]{hyperref}

\usepackage{color}

\newcommand{\CCOC}{$\mathrm{Ca_2CuO_2Cl_2}$}
\newcommand{\vdCCOC}{$\mathrm{Ca_{1.84}CuO_2Cl_2}$}
\newcommand{\NaCCOC}{$\mathrm{Ca_{2-x}Na_xCuO_2Cl_2}$}
\newcommand{\VCCOC}{$\mathrm{Ca_{2-x}CuO_2Cl_2}$}
\newcommand{\LSCO}{$\mathrm{La_{2-x}Sr_xCuO_4}$}
\newcommand{\LSCOopt}{$\mathrm{La_{1.85}Sr_{0.15}CuO_4}$}
\newcommand{\LSCOover}{$\mathrm{La_{1.7}Sr_{0.3}CuO_4}$}
\newcommand{\YBCO}{$\mathrm{YBa_2Cu_3O_{6+\delta}}$} 
\newcommand{\LCO}{$\mathrm{La_2CuO_4}$}

\begin{document}

%\preprint{}
%\date{\today}

%%%%%%%%%%
%%%%%%%%%%
\title{Phonon anomalies with doping in superconducting oxychlorides \texorpdfstring{Ca$_{2-x}$CuO$_2$Cl$_2$}{Ca2-xCuO2Cl2}}
%%%%%%%%%%
%%%%%%%%%%

\author{Blair W. Lebert}
\affiliation{IMPMC, UMR CNRS 7590, Sorbonne Universit\'es-UPMC University Paris 06, MNHN, IRD, 4 Place Jussieu, F-75005 Paris, France}

\author{Hajime Yamamoto}
\author{Masaki Azuma}
\affiliation{Materials and Structures Laboratory, Tokyo Institute of Technology, 4259 Nagatsuta, Midori-ku, Yokohama, 226-8503, Japan}

\author{Rolf Heid}
\affiliation{Institute for Solid State Physics, Karlsruhe Institute of Technology, D-76021 Karlsruhe, Germany}

\author{Satoshi Tsutsui}
\author{Hiroshi Uchiyama}
\affiliation{Japan Synchrotron Radiation Research Institute (JASRI), SPring-8, 1-1-1 Kouto, Sayo, Hyogo 679-5198, Japan}

\author{Alfred Q.R. Baron}
\affiliation{Materials Dynamics Laboratory, RIKEN SPring-8 Center, RIKEN, 1-1-1 Kouto, Sayo Hyogo 679-5148. }

\author{Beno\^it  Baptiste}
\affiliation{IMPMC, UMR CNRS 7590, Sorbonne Universit\'es-UPMC University Paris 06, MNHN, IRD, 4 Place Jussieu, F-75005 Paris, France}

\author{Matteo d'Astuto}\email{matteo.dastuto@neel.cnrs.fr}
\affiliation{IMPMC, UMR CNRS 7590, Sorbonne Universit\'es-UPMC University Paris 06, MNHN, IRD, 4 Place Jussieu, F-75005 Paris, France}
\affiliation{Institut NEEL CNRS/UGA UPR2940, 25 rue des Martyrs BP 166, 38042 Grenoble cedex 9}

%%%%%%%%%%
%%%%%%%%%%
\begin{abstract}
We measure the dispersion of the Cu-O bond-stretching phonon mode in the high-temperature superconducting parent compound \CCOC{}. Our density functional theory calculations predict a cosine-shaped bending of the dispersion along both the ($\xi$00) and ($\xi\xi$0) directions, while comparison with previous results on \vdCCOC{} show it only along ($\xi$00), suggesting an anisotropic effect which is not reproduced in calculation at optimal doping. 
Comparison with isostructural \LSCO{} suggests that these calculations reproduce well the overdoped regime, however they overestimate the doping effect on the Cu-O bond-stretching mode at optimal doping. 
\end{abstract}
%%%%%%%%%%
%%%%%%%%%%

%74.72.Gh Hole-doped Cuprate superconductors; 
%63.20.D- Phonon states and bands, normal modes, and phonon dispersion 
%63.20.kd Phonon-electron interactions
%78.70.Ck X-ray scattering 


\maketitle

%%%%%%%%%%%%%%%%%%%%%%%%%%%%%%%%%%
%%%%%%%%%% Introduction %%%%%%%%%%
%%%%%%%%%%%%%%%%%%%%%%%%%%%%%%%%%%
The role of electron-phonon coupling in high-temperature superconducting (HTS) cuprates has been debated since their discovery \cite{bm}. Although the general belief is that coupling with phonons is not the main mechanism driving Cooper pair formation in HTS cuprates \cite{nat-phys-bonn-rv}, their role is still not completely understood. For example, the electron-phonon coupling exhibits anomalous doping dependence with a very large oxygen isotope effect close to 1/8 doping \cite{Crawford-isoLaCuO,Chen3732}. A decade ago, the debate around the role of electron-phonon coupling was revived by the observation of a strong kink in the electronic band dispersion measured by angle-resolved photoemission spectroscopy (ARPES) \cite{lanzara-kink} which was thought to originate from phonon interactions. The Cu-O bond stretching phonon \cite{pintrev1,mcqueeney,dastuto-ncco,uchiyama,fukuda,reznik,reznik-rev}, which softens with doping, is the most likely candidate for this interaction \cite{jeff-bisco}. Subsequent density functional theory (DFT) calculations \cite{bohnen,giustino,PhysRevLett.100.137001}, could explain rather well the phonon softening despite small electron-phonon coupling, but they could not explain the large ARPES kink. However, it was suggested that large couplings may still exist due to many-body effects in the presence of strong electron-electron correlations \cite{rosch-gunn} which are not captured by these DFT calculations.

\begin{figure}[htb]
\includegraphics[width=0.45\linewidth]{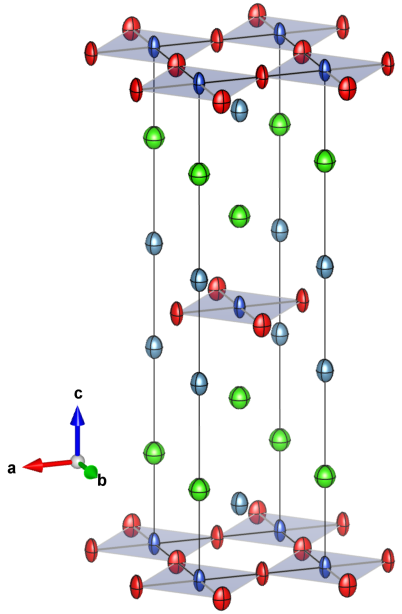}
\caption{\label{ccoc structure} (Color online) The crystallographic unit cell of \CCOC{} with Ca as cyan, Cu as blue, O as red, and Cl as green \cite{VESTA}. The square coordination of Cu with its four nearest-neighbor O ions in the CuO$_2$ planes is shown. The Cl ions are located at the apical sites below/above the Cu ions. The atomic coordinates and displacement ellipsoids are from single-crystal diffraction detailed in Ref.~\onlinecite{baptiste-ccoc-cif}.}
\end{figure}

In this Letter, we present inelastic x-ray scattering (IXS) measurements of the parent compound \CCOC{} \cite{hiroi,kohsaka-jacs}. We demonstrate that doping induces a softening of the Cu-O bond-stretching phonon by comparing with previous reports \cite{dastuto-ccocoPRB} on the vacancy-doped compound \vdCCOC{} \cite{yamada-ccoc}, which is near optimal doping. This result is consistent with the above cited reports of doping-induced softening in other HTS cuprates. The softening however is anisotropic which disagrees with our DFT calculations. We show by comparison with \LSCO{} \cite{pint-lascuo-od}, since \VCCOC{} cannot be overdoped, that DFT calculations actually reproduce the strongly overdoped region in HTS cuprates. The failure of the DFT calculations to reproduce this important phonon mode near optimal doping naturally explains its inability to reproduce the observed large ARPES kink. In the future our results on \CCOC{}, coupled with previous reports on \vdCCOC{} \cite{dastuto-ccocoPRB}, may help bridge this gap between theory and experiment in the HTS cuprates. The \CCOC{} system is ideally suited to advanced many-body calculations trying to capture the predicted larger electron-phonon coupling due to electronic correlations because of its light elements and simple structure \cite{Foyevtsova2014a,wagner-qmc-oxychlo}.

%%%%%%%%%%%%%%%%%%%%%%%%%%%%%
%%%%%%%%%% Methods %%%%%%%%%%
%%%%%%%%%%%%%%%%%%%%%%%%%%%%%
Single crystals of \CCOC{} were grown by the flux method as described in Ref.~\onlinecite{baptiste-ccoc-cif}. The phonons were measured using inelastic x-ray scattering (IXS) at the BL35XU beamline of SPring-8 \cite{BARON2000461}. Grease/oil was used to protect the hygroscopic samples from air and to mount them on copper sample holders in a cryostat. The cryostat was used just for its vacuum to protect the samples and minimize air scattering. However, the measurements were taken at room temperature which is possible with \CCOC{} because the lower energy modes are weaker due to the low $Z$ atoms and therefore do not wash out the higher energy modes \cite{dastuto-ccocoPRB}. The main monochromator was set to the Si(999) Bragg reflection giving a wavelength of 0.6968 \AA{} (17.7935~eV) and the beam size at the sample was 0.09~$\times$~0.09~mm$^2$ FWHM (see Ref.~\cite{BARON2000461} for details). The angular width of the (400) Bragg reflection rocking curve from the samples was $\approx0.3^{\circ}$ FWHM.

DFT calculations of the phonon dispersion for \CCOC{} and \LCO{} were carried out using the linear response or density-functional perturbation theory implemented in the framework of the mixed-basis pseudopotential method \cite{Heid}. The lattice structure of \CCOC{} was fully relaxed prior to the phonon calculations. In the case of \LCO{}, we used the experimental lattice constants of \LSCO{} with x=0.3 \cite{Rad} and only
relaxed the internal structural parameters. In both systems we have used the stoichiometry of the undoped parent compounds, however the present LDA calculations are unable to describe the charge-transfer insulating ground state and instead predict a metallic state. Thus the calculated phonon dispersions are more representative of the doped compounds. Shell calculations were based on a common interatomic potential model for cuprates \cite{chaplot} and adapted to \vdCCOC{} in a previous work  \cite{dastuto-ccocoPRB}.

%%%%%%%%%%%%%%%%%%%%%%%%%%%%%
%%%%%%%%%% Results %%%%%%%%%%
%%%%%%%%%%%%%%%%%%%%%%%%%%%%%

\begin{figure}[t]
\includegraphics[width=\linewidth]{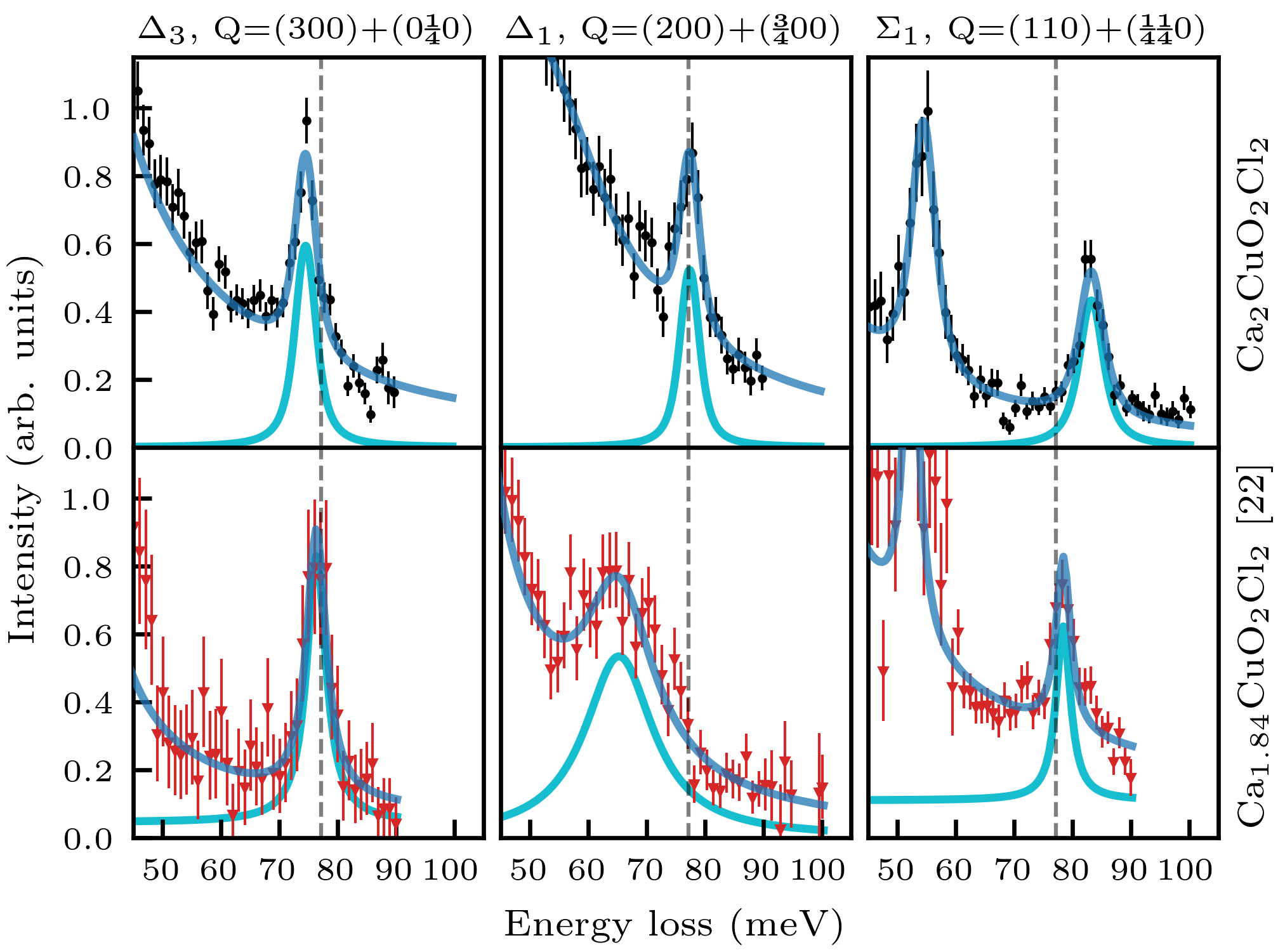}
\caption{\label{example-raw} (Color online) Representative inelastic x-ray scattering spectra of \VCCOC{}. Measurements on \CCOC{} from this work, shown in the top row, are compared to previous reports on \vdCCOC{} \cite{dastuto-ccocoPRB} shown on the bottom row. Each column shows spectra taken at the propagation vector corresponding to the middle of the Brillouin zone along three different in-plane symmetry lines. Blue lines are fit to the entire data, while the cyan line highlights the contribution from the Cu-O bond stretching mode. The vertical dashed gray line is the position at $\Gamma$ from infrared absorption \cite{ccoc-ir}.
}
\end{figure}

\begin{figure*}[t]
\includegraphics[width=\linewidth]{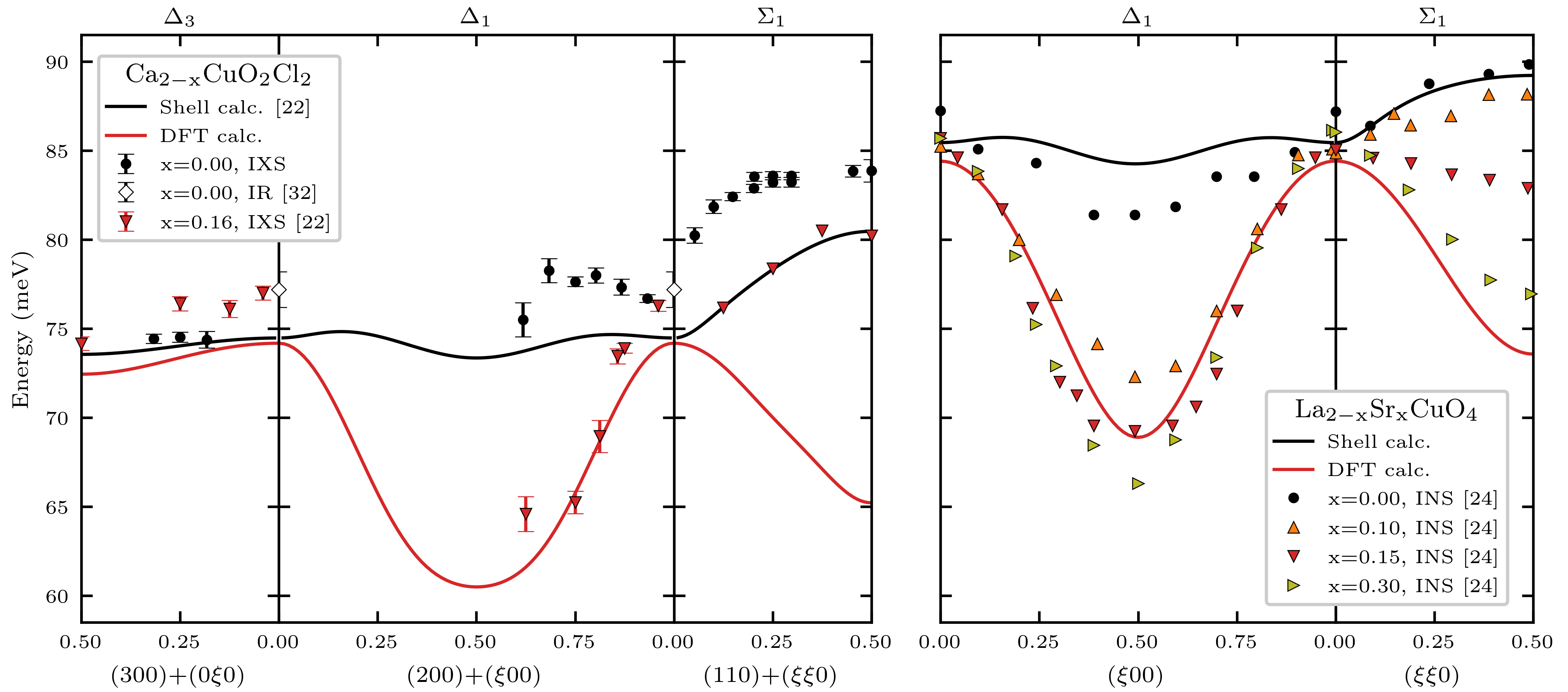}
\caption{\label{disp} (Color online) Dispersion of Cu-O bonding-stretching phonon in \VCCOC{} (left) and \LSCO{} (right). The inelastic x-ray scattering (IXS) on \CCOC{} from this work are plotted along with: infrared absorption (IR) taken at $\Gamma$ on \CCOC{} \cite{ccoc-ir}, IXS on \vdCCOC{} \cite{dastuto-ccocoPRB}, and inelastic neutron scattering (INS) on \LSCO{} (x=\{0, 0.1, 0.15, 0.3\}) \cite{pint-lascuo-od}. Shell calculations are shown as black lines and density functional theory (DFT) calculations are shown as red lines.
}
\end{figure*}

In Fig.~\ref{example-raw} we show representative IXS spectra for \CCOC{} (top) and \vdCCOC{} (bottom, from Ref.~\onlinecite{dastuto-ccocoPRB}) at the midpoints of the three symmetry lines we explored: $\Delta_1$, longitudinal along ($\xi$00); $\Delta_3$, transverse along (0$\xi$0); and $\Sigma_1$, longitudinal along ($\xi\xi$0). The blue lines are a fit of the entire spectra consisting of Lorentzian functions convoluted with the instrumental function, while the cyan lines shown the Cu-O bond stretching phonon contribution. 

Our results are summarized in Fig.~\ref{disp} where we compare the measured and calculated Cu-O bond-stretching phonon dispersion of \VCCOC{} (left) and \LSCO{} (right). Our dispersion from IXS on \CCOC{}, shell calculations on \LSCO{}, and DFT calculations on \CCOC{} and \LCO{} are complemented by infrared absorption measurements on \CCOC{} \cite{ccoc-ir}, shell calculations on \VCCOC{} \cite{dastuto-ccocoPRB}, and dispersion from inelastic neutron scattering (INS) on \LSCO{} \cite{pint-lascuo-od}.

Our measurements of \CCOC{} confirm that near optimal doping the Cu-O bond-stretching phonon in \VCCOC{} softens along $\Delta_1$, which agrees with previous reports on \LSCO{} and other HTS cuprates  \cite{pintrev1,mcqueeney,dastuto-ncco,uchiyama,fukuda,reznik,reznik-rev}. However, with respect to the undoped compounds, there is an upward dispersion in \CCOC{} unlike the downward dispersion found in \LCO{}. There is also a doping-induced softening along $\Sigma_1$, however the upward dispersion of \CCOC{} persists with doping unlike the downward bending seen in \LSCO{} with doping. We find a strangely fast dispersion along $\Sigma_1$ near the zone center in \CCOC{} which does however decrease upon doping. The measurements along $\Delta_3$ with transverse polarization show no doping dependence, which stresses the fact that only the Cu-O bond stretching mode is softened with doping, as for the other cuprates \cite{pintrev1,dastuto-ncco,reznik-rev}

Our DFT and shell model empirical calculations are shown in Fig.~\ref{disp} as red and black lines respectively. We stress that the DFT calculations are more representative of doped HTS cuprates, despite being performed with an undoped stoichiometry, since they cannot open the charge-transfer gap and instead predict a metallic state. Indeed, we find good agreement between these calculations along $\Delta_1$ for both \vdCCOC{} and \LSCOopt{}. On the other hand, the shell model empirical calculations account for screening and are fit to doped samples, however they are more representative of undoped HTS cuprates as seen in Fig.~\ref{disp}. Except for a slight shift, the higher optic mode along $\Delta_3$ agrees well with shell calculations.

The DFT calculations predict a softening along $\Sigma_1$ similar to that along $\Delta_1$. We found however that near optimal doping DFT calculations fail in both compounds along $\Sigma_1$. The difference near the zone center is small, however it grows larger near the zone boundary since DFT predicts a strong downward dispersion towards the zone boundary. We find the opposite trend in \vdCCOC{} which actually has an upward dispersion, while \LSCOopt{} does disperse downwards but quite weakly. 

The apparent contradiction between theory and experiment is resolved by considering the overdoped HTS cuprates. As shown in the right panel of Fig,~\ref{disp}, the dispersion of \LSCOover{} along both $\Delta_1$ and $\Sigma_1$  agrees much better with DFT calculations. We conclude that standard DFT calculations on HTS cuprates overestimate the doping effects of the Cu-O bond-stretching mode. 

Unfortunately, a similar comparison cannot be made with \CCOC{} since it has never been overdoped, neither with sodium \cite{hiroi,kohsaka-jacs} nor with vacancies \cite{yamada-ccoc}. Nonetheless, our IXS results on \CCOC{} coupled with those on \vdCCOC{} \cite{dastuto-ccocoPRB} provide an experimental test bed for future theoretical calculations trying to improve upon DFT by including correlation effects. In order to minimize relativistic effects, these quantum many-body calculations are mainly done on systems with light atoms \cite{Foyevtsova2014a, Wagner2015}. \VCCOC{} and \NaCCOC{} are the closest examples to such systems among the bulk HTS cuprates. The \CCOC{} system also has the advantage of a simple single-layer quadratic structure without any doping- or temperature-induced structural transitions which can affect phonon mode frequencies. 

The failure of DFT calculations to reproduce the dispersion along $\Sigma_1$ at optimal doping was also found in \YBCO{} \cite{Heid2007} which suggests that this phenomenon is universal in the HTS cuprates. In Ref.~\onlinecite{giustino} the $\Sigma_1$ mode is not shown, but earlier DFT phonon calculations \cite{PhysRevB.59.9278} found a similar difference with experiment. The authors of Ref.~\onlinecite{PhysRevB.59.9278} noted that the difference was not as drastic and temperature-sensitive as previous models which included Jahn-Teller effects, however they did not elaborate further on the actual difference \cite{PhysRevB.45.5633}. Moreover, their calculated frequencies were shifted since they used an idealized tetragonal structure for undoped \LCO{}. On the contrary, our present calculations shown in Fig.~\ref {disp} use the experimental lattice constants for \LSCOover{} and agree with experimental results at $\Gamma$ and along the other branches without an energy shift. The difference between DFT calculations and experiment along $\Sigma_1$ in the underdoped to optimally doped regime can be simulated using phenomenological models of the dipole and charge fluctuations \cite{falter-lsco}, however this technique is not first-principles since it uses a shell-model approach with fitted parameters.

In conclusion, we show that the softening of the Cu-O bond-stretching mode induced by doping in \VCCOC{} is anisotropic near optimal doping, with a marked difference along $\Delta_1$ and $\Sigma_1$, \textit{i.e.} full- and half-breathing modes. This is in striking disagreement with DFT calculations which we show actually reproduces the modes in overdoped cuprates, using \LSCO{} as an example. This in turn could explain a smaller calculated effect on the ARPES extracted self-energy in DFT. 
An anisotropic electron-phonon coupling could be relevant to understanding the physics of cuprate superconductivity as pointed out by \cite{Chen3732}, and oxychloride cuprates are an optimal playground to test advanced many-body calculations trying to capture the effects of electronic correlations.

%%%%%%%%%%%%%%%%%%%%%%%%%%%%%%%%%%%%%
%%%%%%%%%% Acknowledgments %%%%%%%%%%
%%%%%%%%%%%%%%%%%%%%%%%%%%%%%%%%%%%%%
\begin{acknowledgments}
The synchrotron radiation experiments were performed at the BL35XU of SPring-8 with the approval of the Japan Synchrotron Radiation Research Institute (JASRI) (Proposal No. 2015B1720). The authors are very grateful to Lise-Marie Chamoreau for her assistance in sample orientation and acknowledge the use of the  x-ray diffractometer instrument at the ``Plateforme Diffraction'', IPCM, Paris. The authors thank Laura Chaix for critical reading of the manuscript. 
B.W.L acknowledges financial support from the French state funds managed by the ANR within the ``Investissements d'Avenir'' programme under reference  ANR-11-IDEX-0004-02, and within the framework of the Cluster of Excellence MATISSE led by Sorbonne Universit\'{e} as well as from the LLB/SOLEIL PhD fellowship program. 
B.W.L also acknowledges travel within Japan, housing and user fees covered by SPring-8 through the "Budding Researchers Support" program. 
\end{acknowledgments}

%%%%%%%%%%%%%%%%%%%%%%%%%%%%%%%%%%
%%%%%%%%%% Bibliography %%%%%%%%%%
\bibliographystyle{apsrev4-1}

%%%%%%%%%%%%%%%%%%%%%%%%%%%%%%%%%%
%\bibliography{cup-ph-bib}
%\bibliographystyle{apsrev4-1}
\end{document}